\documentclass[twocolumn,showpacs,prb,preprintnumbers,superscriptaddress,amsmath,amssymb]{revtex4}

\usepackage{graphicx}
\usepackage{dcolumn}
\usepackage{bm}

\begin{document}

\preprint{APS/123-QED}

\title{Vortex trapping and expulsion in thin-film YBa$_2$Cu$_3$O$_{7-\delta}$ strips}

\author{K.~H.~Kuit}
\affiliation{Low Temperature Division, Mesa$^+$ Institute for Nanotechnology,
University of Twente, P.O. Box 217, 7500 AE Enschede, The Netherlands}

\author{J.~R.~Kirtley}
\affiliation{Low Temperature Division, Mesa$^+$ Institute for Nanotechnology,
University of Twente, P.O. Box 217, 7500 AE Enschede, The Netherlands}
\affiliation{Department of Applied Physics, Stanford University, Palo Alto, CA, USA}
\affiliation{Department of Microelectronics and Nanoscience, Chalmers University of Technology, S-41296 G{\oe}teborg, Sweden}

\author{W.~van der Veur}
\affiliation{Low Temperature Division, Mesa$^+$ Institute for Nanotechnology,
University of Twente, P.O. Box 217, 7500 AE Enschede, The Netherlands}

\author{C.~G.~Molenaar}
\affiliation{Low Temperature Division, Mesa$^+$ Institute for Nanotechnology,
University of Twente, P.O. Box 217, 7500 AE Enschede, The Netherlands}

\author{F.~J.~G.~Roesthuis}
\affiliation{Low Temperature Division, Mesa$^+$ Institute for Nanotechnology,
University of Twente, P.O. Box 217, 7500 AE Enschede, The Netherlands}

\author{A.~G.~P.~Troeman}
\affiliation{Low Temperature Division, Mesa$^+$ Institute for Nanotechnology,
University of Twente, P.O. Box 217, 7500 AE Enschede, The Netherlands}

\author{J.~R.~Clem}
\affiliation{Ames Laboratory-DOE and Department of Physics and Astronomy,
Iowa State University, Ames, Iowa 50011, USA}

\author{H.~Hilgenkamp}
\affiliation{Low Temperature Division, Mesa$^+$ Institute for Nanotechnology,
University of Twente, P.O. Box 217, 7500 AE Enschede, The Netherlands}

\author{H.~Rogalla}
\affiliation{Low Temperature Division, Mesa$^+$ Institute for Nanotechnology,
University of Twente, P.O. Box 217, 7500 AE Enschede, The Netherlands}

\author{J.~Flokstra}
\affiliation{Low Temperature Division, Mesa$^+$ Institute for Nanotechnology,
University of Twente, P.O. Box 217, 7500 AE Enschede, The Netherlands}

\date{\today}

\begin{abstract}
A scanning SQUID microscope was used to image vortex trapping
as a function of the magnetic induction during cooling
in thin-film YBa$_2$Cu$_3$O$_{7-\delta}$ (YBCO) strips for
strip widths $W$ from 2 to 50 $\mu$m.
We found that vortices were excluded from the strips when the induction $B_{a}$ was
below a critical induction $B_c$.
We present a simple model for the vortex exclusion process which takes into account the
vortex - antivortex pair production energy as well as the vortex Meissner and self-energies.
This model predicts that the real density $n$ of trapped vortices is given
by $ n = (B_a - B_K)/\Phi_0 $ with $B_K = 1.65 \Phi_0/W^2$ and $\Phi_0=h/2e$ the
superconducting flux quantum. This prediction is
in good agreement with our experiments on
YBCO, as well as with previous experiments on thin-film strips of niobium.
We also report on the positions of the trapped vortices. We found that at low densities
the vortices were trapped in a single row near the centers of the strips, with the relative
intervortex spacing
distribution width decreasing as the vortex density increased, a sign of longitudinal ordering.
The critical induction
for two rows forming in the 35 $\mu$m wide strip was $(2.89+1.91-0.93) B_c$, consistent
with a numerical prediction.
\end{abstract}

\maketitle

\section{\label{sec:intro}Introduction}
In principle, when a parallel magnetic field is applied to an infinitely long, defect-free superconducting
cylinder, all magnetic flux should be expelled as the temperature $T$ is lowered through the
superconducting transition temperature $T_c$, provided that the applied magnetic field is below either the
critical field $H_c(T)$ for a type-I superconductor, or the lower critical
field $H_{c1}(T)$ for a type-II superconductor.\cite{tinkham96} In practice, real samples have finite size
and often contain defects, which can pin magnetic flux. Moreover, nonellipsoidal samples, even those not
containing defects, naturally possess geometric energy barriers that can trap magnetic flux during the
cooling process.  Pinned or trapped vortices are nearly always observed in thin-film type-II
superconductors, even when cooled in relatively low magnetic fields.
In general, this can be attributed both to pinning of vortices by, for example, defects and grain boundaries,
and to trapping by the geometric energy barriers.  Understanding such pinning and trapping effects is
important for
superconducting electronics applications.

The present work is motivated by applications of high-$T_c$ superconducting sensors such as
SQUIDs\cite{5553153} and hybrid magnetometers based on high-$T_c$ flux concentrators.\cite{7287036}
These sensors are used
in a broad field of applications, such as geophysical research\cite{7715464} and biomagnetism.\cite{7448298}
The sensitivity of these sensors is limited by $1/f$ noise in an unshielded environment.
The dominant source of this noise is the movement of vortices trapped in the sensor.
This noise
can be eliminated by dividing the high-$T_c$ body into thin strips.\cite{6356793,5553153} The strips
have a certain critical induction below which no vortex trapping occurs, resulting in an ambient field range
in which these sensors can be effectively operated.
We investigated vortex trapping in thin-film YBCO strips in
order to incorporate the results in a hybrid magnetometer based on a YBCO
ring tightly coupled to, for example, a GMR (giant magneto resistance) or Hall sensor.

Models for the critical induction of thin-film strips have been proposed by Clem\cite{Clem} and
Likharev.\cite{Likharev1972} Indirect experimental testing of these models was done
by observing noise in high-$T_c$ SQUIDs as a function of
strip width and induction.\cite{5479111,6356793,5553153} The induction mentioned here is the magnetic induction during cooling, which is the notation throughout this paper.
More direct experimental verification of these models was presented by
Stan et al.\cite{7890333} using scanning Hall probe microscopy (SHPM) on Nb strips.
Both experiment and theory found that the critical induction varied roughly like $1/W^{2}$.
However, the experimental\cite{7890333} and theoretical\cite{Clem,Likharev1972}
pre-factors multiplying this $1/W^{2}$ dependence differed significantly. In this
paper we propose a model for vortex trapping in narrow superconducting strips
which takes into account the role of thermally generated vortex-antivortex pairs.

To test this model we performed scanning SQUID microscopy (SSM)\cite{5244937}
measurements on thin-film YBCO strips. We found excellent agreement
between the dependence of critical induction on strip width and the present model
for both our experiments on YBCO and for the previous work on Nb.
In agreement with this previous work and as predicted by the present model, we found that in YBCO the number of vortices increased for inductions above the critical induction linearly with the difference
between the applied induction and the critical induction.
In a follow-up to the paper of Stan et al., Bronson et al.\cite{8909005} presented
numerical simulations for the vortex distribution in narrow strips. These simulations
showed that for inductions just above the critical induction the vortices are trapped in the centers
of the strips. For higher inductions the vortices formed more complex ordered patterns, first in two
parallel rows, then for higher inductions in larger numbers of parallel rows. We performed statistical
analysis of the vortex distribution in our measurements and found agreement with this model.

\section{\label{sec:Theory} Theory of vortex trapping in a thin film strip}

Whether or not a vortex gets trapped in a strip is determined by the Gibbs free energy. This energy
exhibits a dip in the center of a superconducting strip
for applied inductions above a certain critical value. This dip gives rise to an energy
barrier for the escape of the vortex. The models proposed by Clem\cite{Clem}
and Likharev\cite{Likharev1972} differ from the present model only
in the minimum height of the energy barrier required to trap vortices.

\subsection{The Gibbs free energy of a vortex in a strip}

Consider a long, narrow, and thin superconducting strip of width $W$ in an applied magnetic induction
$B_a$. The vortex trapping
process occurs sufficiently close to the superconducting transition temperature that the Pearl
length $\Lambda = 2 \lambda^2/d$, with $\lambda$ the London penetration depth and $d$ the film thickness,
is larger than
$W$.
In this limit there is little shielding
of an externally applied magnetic induction $B_a$. The
resultant superconducting currents
in the strip can be calculated using the fluxoid quantization condition:\cite{Tinkham}
\begin{equation}
\int \int \vec B \cdot d \vec S + \mu_0 \lambda^2 \oint \vec J_s \cdot d \vec s =
N \Phi_0. \label{Tinkham}
\end{equation}
In this equation the first integration is over a closed surface $S$ within the superconductor,
the second is over a
closed contour surrounding $S$, $\vec B$ is the magnetic induction,
$\vec J_s$ is the supercurrent density,
and $N$ is an integer. SI units are
used throughout this paper. If we take the strip with its long dimension in the $y$ direction,
with edges at $x = 0$ and $x = W$, and an applied induction perpendicular to the strip in the $z$
direction, a square closed contour can be drawn with sides at $y = \pm l/2$ and $x = W/2 \pm \Delta x$.
If we assume uniform densities $n_v$ and $n_a$ of vortices and antivortices in the film,
with $n = n_v - n_a$ being the excess density of vortices over antivortices, the first
integral in Eq.\ (\ref{Tinkham})
becomes $2 B_a \Delta x l$, the second becomes $2J_s l$, $N = 2nl \Delta x$, and the supercurrent induced
in response to the applied induction is:
\begin{equation}
J_{y} = -\frac{1}{\mu_0 \lambda^2}\left( B_a - n\Phi_0 \right) \left( x - W/2 \right) \label{superc}.
\end{equation}
The assumption of a uniform density of vortices is good at high trapping densities, and at zero density,
but is incorrect at low densities, as we shall discuss later.
Equation (\ref{superc}) differs from the expression given in Ref.\ \onlinecite{Clem} by the term $-n\Phi_0$:
As vortices are nucleated in the film, they reduce the screening
currents induced in the film by the applied induction. The equation derived in Ref.\ \onlinecite{Clem} for the Gibbs free
energy of an isolated vortex (upper sign) or an antivortex (lower sign) at a position $x$ inside
the strip is then slightly modified as: \cite{4739733,6085150}
\begin{eqnarray}
G(x) = && \frac{\Phi_0^2}{2\pi\mu_0\Lambda}\ln\left[\frac{\alpha W}{\xi}\sin\left(\frac{\pi x }{W}\right)\right]
\nonumber \\
&& \mp \frac{\Phi_0 \left(B_a - n\Phi_0 \right)}{\mu_0 \Lambda}x\left(W-x\right).\label{Genergy}
\end{eqnarray}
The Gibbs free energy consists
of two terms. The first term, which is independent of the applied magnetic induction $B_a$, is
calculated to logarithmic accuracy, as it includes only the kinetic energy of the supercurrents,
and it is equal to $\Phi_0 I_{circ}/2$, where $I_{circ}$ is the supercurrent circulating around
the vortex.  This term, which has a dome shape and
decreases monotonically to zero as the vortex reaches a distance $\xi/2$ from the edges
of the strip, is also equal to the work that must be done to move the vortex from its initial
position at $x = \xi/2$ or $x = W-\xi/2$ to its final position at $x$ against the Lorentz forces
of attraction between the vortex and an infinite set of negative image vortices at $-x+2mW$,
$m = 0, \pm 1, \pm 2, ...$ .  Here  $\xi$ is the coherence length, which is assumed to obey $\xi \ll W$.
We also assume that the vortex core radius is $\xi$, such that the constant $\alpha=2/\pi$ as in Ref.\ \onlinecite{Clem}.
Other values of $\alpha$, such as  $1/\pi$ as in Ref.\ \onlinecite{6085150}, or  $1/4$ as in Ref.\ \onlinecite{Likharev1972}
correspond to different assumptions regarding the core size.
The
second term in Eq.\ (\ref{Genergy}) is the interaction energy between a vortex (upper sign)
[or an antivortex (lower sign)] and the screening currents induced by the
external magnetic induction.  It is the negative of the work required to bring a vortex (or antivortex)
in from the edge against the Lorentz force due to the induced supercurrent given in Eq.\ (\ref{superc}).
The upper sign in Eq.\ (\ref{Genergy}) corresponds to the fact that $J_{ay}$ tends to drive  vortices
into the film, and the lower sign indicates that antivortices are driven out.  When $B_a$ is
sufficiently large, this term makes a minimum in $G(x)$ in which vortices can be trapped. For wider
strips this minimum occurs at lower values of the induction.

\subsection{\label{sec:excist}Previous models for the critical induction}

There are two existing models which predict the critical induction for vortex trapping when the
applied perpendicular magnetic induction is small ($B_a \sim \Phi_0/W^2$). In these models
the Gibbs free energy from Eq.\ (\ref{Genergy}) is used in the limit of $n \rightarrow 0$.
The critical induction model by Likharev\cite{Likharev1972} states that in order
to trap a vortex in a strip the vortex should be absolutely stable. This happens when the Gibbs free energy
in the middle of the strips equals zero and leads to
\begin{equation}
B_L = \frac{2\Phi_0}{\pi W^2}\ln\left(\frac{\alpha W}{\xi} \right),\label{likharev}
\end{equation}
where $\alpha$ is the constant in Eq.\ (\ref{Genergy}).

Another model for the critical induction is proposed by Clem,\cite{Clem} who considers a metastable condition.
In this view vortex trapping will occur when the applied magnetic induction is just large enough
to cause a minimum in the Gibbs free energy at the center of the strip, $d^2 G(W/2)/dx^2 = 0$,
leading to
\begin{equation}
B_0 = \frac{\pi \Phi_0}{4 W^2}. \label{Clem}
\end{equation}

\subsection{\label{sec:our}Our model for the critical induction}

The model proposed here is intermediate between the models presented in Ref.\ \onlinecite{Clem} and \onlinecite{Likharev1972}.
As the strip is cooled just below
the superconducting transition temperature $T_c$, thermal fluctuations cause the generation of a
high density of vortex-antivortex pairs. Similar to the processes determining the equilibrium
densities of electrons and holes in semiconductors, the equilibrium densities of vortices and
antivortices very near $T_c$ are determined by a balance between the rate of generation of
vortex-antivortex pairs, the rate of their recombination, and the rates with which vortices
are driven inwards and antivortices are driven outwards by the current $J_y$. Accordingly,
very close to $T_c$, the densities $n_v$ and $n_a$ of vortices and antivortices equilibrate
such that their difference $n = n_v - n_a$ is very nearly equal to $B_a/\phi_0$ and the current
$J_y$ [see Eq.\ (\ref{superc})] is practically zero. When $B_0<B_a<B_L$, it is energetically
unfavorable for vortices and antivortices to be present in the strip, and as the temperature
decreases and the energy scales of the terms in Eq.\ (\ref{Genergy}) increase, the densities
of both vortices and antivortices decrease.  While vortices and antivortices continue to be
thermally generated, the antivortices are quickly driven out of the strip by the combination
of the self-energy and field-interaction energy [note the lower sign in Eq.\ (\ref{Genergy})].
The antivortex density thus becomes much smaller than the vortex density and becomes so small
that the recombination rate is negligible.  The value of $n \approx n_v$ drops below $B_a/\phi_0$.
Although when  $B_0<B_a<B_L$ it is energetically unfavorable for a vortex to be present in the strip,
the vortex's Gibbs free energy has a local minimum at the center of the strip and the vortex must
overcome the energy barrier before it can leave the strip.  Since the energy required to form a
vortex-antivortex pair is given by the pairing energy:\cite{tinkham96}
\begin{equation}
E_{pair} = \frac{\Phi_0^2}{4 \pi\mu_0 \Lambda}\label{pair},
\end{equation}
the vortex-antivortex pair generation rate is given by a pre-factor
times the Arrhenius factor $\exp(-E_{pair}/k_B T)$, where $k_B$ is the Boltzmann constant. The
vortex escape rate is given by an attempt frequency times
a second Arrhenius factor $\exp(-E_B/k_B T)$, where $E_B$ is
the difference in the Gibbs free energy between the local maximum and the minimum in
the center of the strip. Since $E_B$ and $E_{pair}$ have the same temperature dependences (recall
that $1/\Lambda$ is proportional to $T_c-T$), the vortex generation rate and its rate of escape
will be exactly  balanced at all
temperatures (aside from
a logarithmic factor in the ratio of the two pre-factors) when $E_B$ and $E_{pair}$ are equal.
This occurs at a critical magnetic induction $B_K$ which is the solution
of the equation
\begin{equation}
\max\left[G(x) - G(W/2)\right] = E_{pair},
\end{equation}
which leads to the condition
\begin{eqnarray}
&& \max\left\{\ln \left[\sin\left(\frac{\pi x}{W}\right) \right] +
\frac{2 \pi B_a}{\Phi_0}\left[\frac{W^2}{4} - x \left(W-x \right) \right] \right\} \nonumber \\
&& = \frac{1}{2},
\end{eqnarray}
where the maximum value of the left-hand side of the equation is taken with respect to $x$.
This equation can be solved numerically, resulting in
\begin{equation}
B_K = 1.65\frac{\Phi_0}{W^2}.\label{Kirtley}
\end{equation}
It appears to be an interesting numerical coincidence that the solution to this equation
gives a pre-factor  (1.6525) that is only
different from $\pi^2/6$ by about 0.5\%.

As the temperature decreases, $\Lambda$ decreases and becomes much less than $W$.
This means that once the vortices are trapped in the local minimum and are clustered
around the middle of the strip, the potential well in which they sit changes shape.
Recall that the calculation of both terms in Eq.\ (\ref{Genergy}) assume that $\Lambda$ is larger than $W$.
It would not even be possible to magnetically image the vortices in the vicinity of the
freeze-in temperature because the local field produced by each vortex is then so spread out.
However, as the temperature decreases, the number of trapped vortices per unit length remains
fixed and the applied field remains constant.  For $n >> 1/W^2$, the distribution of vortices
(averaging over the intervortex spacing) takes on a dome-like shape, and vortex-free zones
appear at the edges of the strip.  The $z$ component of the local magnetic induction should
then be described by the equations given in Sec.\ 2.1 of Ref.\ \onlinecite{brojeny2002}.
The field distribution in a single strip containing a central vortex dome in
which the current density is zero is closely related to the field distribution of a pair of
parallel coplanar strips with a gap between them.\cite{benkraouda1996,benkraouda1998}

\subsection{Behavior above the critical induction}

\begin{figure}[h!]
\includegraphics[width=9cm]{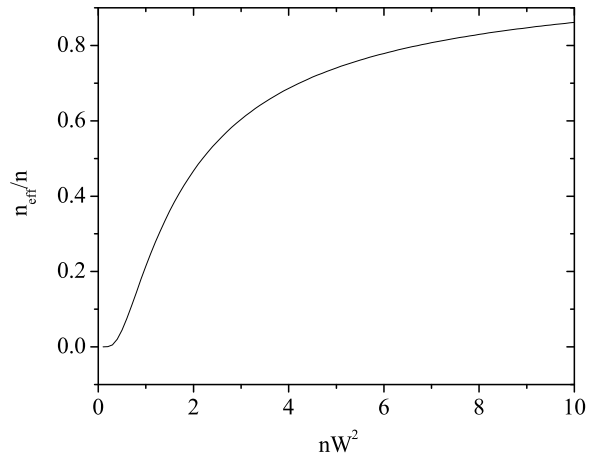}
\caption{\label{fig:neff} Normalized effective vortex density $n_{eff}/n$ vs $nW^2$ from
Eq.\ (\ref{neff}).}
\end{figure}

Because in the present model the screening-current density [Eq.\ (\ref{superc})] and the Gibbs
free energy [Eq.\ (\ref{Genergy})] depend on $n$, the areal density of vortices (when no antivortices are present), we can expect that for
applied inductions $B_a$ well above the critical induction $B_K$ the
balance between the rates of vortex generation and escape occurs when
\begin{equation}
B_a -n\Phi_0 = B_K = 1.65 \frac{\Phi_0}{W^2},
\label{largen}\end{equation}
which can be inverted to give the density $n$ of trapped vortices as a function of
applied induction,
\begin{equation}
n = \frac{B_a - B_K}{\Phi_0}.\label{numv}
\end{equation}
However, for $B_a$ just above $B_K$, where $n \ll B_K/\Phi_0$, we need to take into account the interactions between
vortices more carefully.  As shown by Kogan, \cite{9359454} the interaction energy between a vortex at $(x,0)$ and
another at $(x_i,y_i)$ in a strip of width $W > \Lambda = 2 \lambda^2/d$ and thickness $d$ is \begin{equation}
\epsilon_{int}(x)=\frac{\Phi_0^2}{2\pi\mu_0\Lambda}\ln\Big[\frac{\cosh(\bar y_i)-\cos(\bar x+\bar x_i)}{\cosh(\bar y_i)-\cos(\bar x-\bar x_i)}\Big],\label{epsint}
\end{equation}
where $\bar x = \pi x/W$,  $\bar x_i = \pi x_i/W$,  and $\bar y_i = \pi y_i/W$. To obtain the interaction energy of
the vortex at $(x,0)$ with all the vortices when the vortex density is very large ($n \ll 1/W^2$), it makes sense to
convert the sum over all $i$ to an integral, assuming a uniform density $n$ over the strip width. The integral can be
evaluated analytically, and the result is exactly equal to the term associated with $n\Phi_0$ in Eq.\ (\ref{Genergy}).
However, when the average vortex density is small, we can obtain an approximation to the interaction energy of the
vortex at $(x,0)$ with all other  vortices by again converting the sum to an integral, assuming uniform density of
vortices $n$ over the strip width but only for $|y_i| > 1/2nW$.  In other words, we exclude from the integral a rectangular
region of height $h$ and width $W$  around the origin associated with one vortex, where $n = 1/Wh$.  After changing variables
of integration, the resulting interaction energy $U_{int}(W/2)$ at  the center of the strip can be expressed as
\begin{equation}U_{int}(W/2) = \frac{4n\Phi_0^2W^2}{\pi^3\mu_0\Lambda}\int_0^{\frac{\pi}{2}} d\theta \int_0^{\phi_n} d\phi \frac{\tanh^{-1}(\sin\phi\cos\theta)}{\sin\phi},\label{Uint}
\end{equation}
 where $\phi_n = \sin^{-1}[1/\cosh(\pi/2nW^2)]$.  For a given vortex density $n$, the interaction energy $U_{int}(W/2)$ is less than the value that would be expected from the second term in Eq.\ (\ref{Genergy}).  We may define an effective vortex density $n_{eff}$ by equating $U_{int}(W/2)$ to $n_{eff}\Phi_0^2W^2/4\mu_0\Lambda$, where
\begin{equation}
\frac{n_{eff}}{n}=\frac{16}{\pi^3}\int_0^{\frac{\pi}{2}} d\theta \int_0^{\phi_n} d\phi \frac{\tanh^{-1}(\sin\phi\cos\theta)}{\sin\phi},\label{neff}
\end{equation}
which is plotted in Fig. \ref{fig:neff}.
Since numerical calculations of $U_{int}(x)$ as approximated above show that
\begin{equation}U_{int}\approx \frac{n_{eff}\Phi_0^2}{\mu_0\Lambda}x(W-x),
\end{equation}
we can use the argument that led to Eq.\ (\ref{largen}) to state that the balance between vortex generation and escape occurs when
\begin{equation}
B_a -n_{eff}\Phi_0 = B_K = 1.65 \frac{\Phi_0}{W^2}.
\label{smalln}
\end{equation}
The value of $n$ corresponding to each value of $B_a$ then can be obtained from Eq. (\ref{neff}) or Fig. \ref{fig:neff}. \section{\label{sec:meas}Measurements on YBCO strips}

We performed SSM\cite{5244937} measurements on YBCO
strips. A sample was prepared on a SrTiO$_3$ substrate with a pulsed laser deposited 200 nm
thin film of YBCO. The sample was structured into strips varying from
2-50 $\mu$m in width by Ar ion etching. The SSM, in which the sample was cooled, was placed in a liquid helium
bath cryostat with three layers of $\mu$-metal shielding. The SQUID used in the SSM had a pickup
loop which was defined by focused ion beam milling and had an effective area of 10-15 $\mu$m$^2$
during imaging. A magnetic induction perpendicular to the sample was produced by a solenoid coil which
was placed around the sample and SQUID. After the desired magnetic induction was applied the sample was cooled
to 4.2 K and the sample was scanned. Many different field values have been applied to the sample during cooling to determine the
critical induction for the various strip widths.
The sample was warmed up to well above $T_c$ between different cooling cycles.

\begin{figure}[ht]
\includegraphics[width=8.5cm]{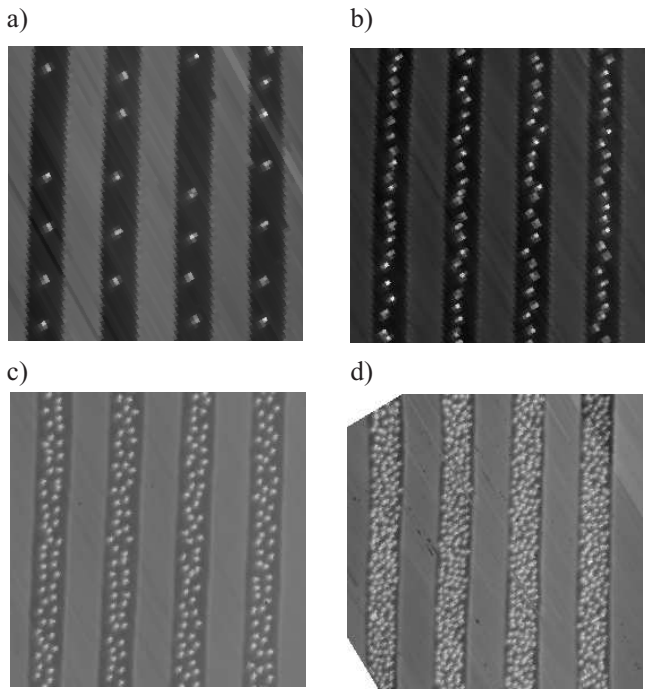}
\caption{\label{fig:SSM} Scanning SQUID microscope images of 35$\mu$m wide YBCO strips
cooled in magnetic inductions of a) 5 $\mu$T, b) 10 $\mu$T, c) 20 $\mu$T, and d) 50 $\mu$T.}
\end{figure}

In Fig. \ref{fig:SSM} SSM images are displayed
of 35 $\mu$m wide strips for several inductions from 5 to 50 $\mu$T. The
strips in these images are darker than their surroundings because of a change in the inductance
of the SQUID sensor as it passed over the superconducting strip.
The bright dots are trapped vortices.
As the inductions increased the vortex density also increased until
it became difficult to distinguish one vortex from the other (Fig. \ref{fig:SSM}d). In Fig. \ref{fig:SSM}a
and Fig. \ref{fig:SSM}b it is clear that at low trapped vortex densities
the vortices tended to form one single row in the center of the strip
where the energy is lowest. In Fig. \ref{fig:SSM}c two parallel lines have been formed, but with
some disorder.

\subsection{\label{sec:bcvsw}Critical induction vs. strip width}

\begin{figure}[h]
\includegraphics[width=9cm]{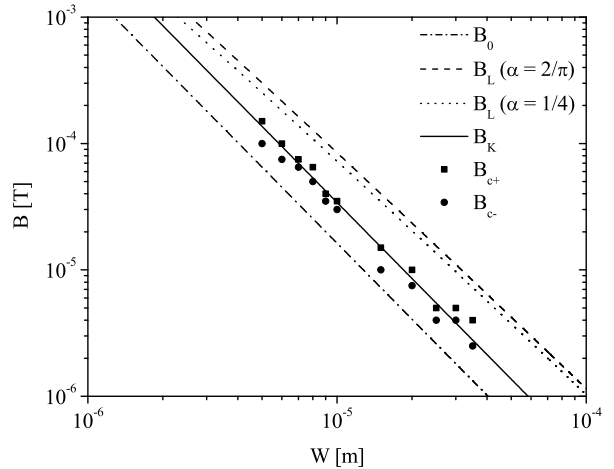}
\caption{\label{fig:bc} Critical inductions for vortex trapping as a function of strip width.
The squares represent $B_{c+}$, the lowest inductions in which trapped vortices were observed, and
the dots
are $B_{c-}$, the highest inductions in which trapped vortices were not observed. The dashed-dotted
line
is the metastable critical induction $B_{0}$ [Eq.\ (\ref{Clem})], the short-dashed
and long-dashed lines are $B_L$ [Eq.\ (\ref{likharev})],
the absolute stability critical inductions calculated at a depinning temperature $T_{dp}=0.98T_c$,
with the constant $\alpha=2/\pi$ \cite{Clem}
or $\alpha=1/4$. \cite{Likharev1972} The solid line is $B_K$ [Eq.\ (\ref{Kirtley})].}
\end{figure}

The results of the measurements of the critical induction vs. strip width are displayed in Fig.\ \ref{fig:bc}
together with the various models. The measurements were performed on strips varying from 6-35 $\mu$m in
width. Measurements on strips narrower than 6 $\mu$m were unreliable because
the critical induction was high enough to degrade the SQUID operation. The critical induction for
40 and 50 $\mu$m wide strips was smaller than the uncertainty
in the applied induction.

There are two data points in Fig.\ \ref{fig:bc} for each strip width:
The upper point indicates the lowest induction at which vortices were observed
trapped in the strip, and the lower point indicates the highest induction at which
vortices were not observed. This provides an upper and lower bound for the actual critical induction.
It was apparent from this log-log plot that the critical induction depended on strip width as a power law.
The best chi-square fit of the experimental data to the two parameter power law $B_c = a\Phi_0/W^p$ yielded
$a=1.96+0.13-0.15$ and $p=1.98\pm0.03$, taking a doubling of the best-fit chi-square as the
uncertainty criterion. If we set the exponent to be $p=2$ ($B_c=a\Phi_0/W^2$), a one parameter fit yielded
$a=1.55\pm0.27$. This is to be compared with $a=1.65$ for the present model [Eq.\ (\ref{Kirtley})], plotted
as $B_K$ in Fig.\ \ref{fig:bc}. It should be emphasized that there were no fitting parameters in
plotting $B_K$.

\begin{figure}[ht]
\includegraphics[width=9cm]{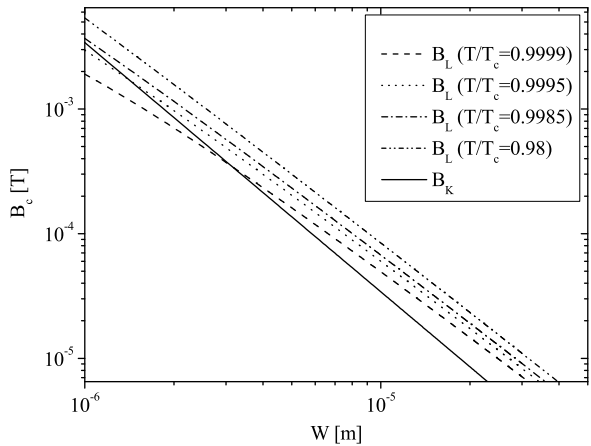}
\caption{\label{fig:tdep} Variation of the prediction of Eq.\ (\ref{likharev}) (using $\alpha=2/\pi$) for
the vortex exclusion critical induction on
depinning temperature (dashed lines). The solid line is $B_K$ [Eq.\ (\ref{Kirtley})].}
\end{figure}

Comparison of the experiment with
the models of Eqs.\ (\ref{Clem}) and (\ref{Kirtley}) is straightforward, since they
are only dependent on the strip width. In order to
evaluate Eq.\ (\ref{likharev}) one must make an estimate of the temperature at which vortex freezout
occurs because of the temperature dependence of $\xi$.
The depinning temperature $T/T_{c}=0.98$  used in Fig.\ \ref{fig:bc}
for both $B_L$ curves
was calculated
by Maurer et al.\cite{5482985} for YBCO.
In addition we used $\xi_{YBCO}(0)=3$ nm, a critical temperature
of T$_c$ = 93 K, and the two-fluid expression for the temperature dependence of the coherence length,
resulting in $\xi(T_{dp}) = 10.39$ nm.
To the best of our knowledge the depinning temperature of YBCO has never
been determined experimentally. Analysis of Eq. \ref{likharev} shows that a $T_{dp}$ closer to $T_c$
could give better agreement between theory and experiment for some strip widths. However
the difference in slopes between theory and experiment becomes larger for higher $T_{dp}$, making
it appear unlikely that this is the correct model for our results.
The dependence of the Likharev model predictions on $T_{dp}$
is displayed in Fig.\ \ref{fig:tdep} for $\alpha=2/\pi$.
For lower $T/T_c$
ratios the curve moves further away from experiment.

\begin{figure}[h]
\includegraphics[width=9cm]{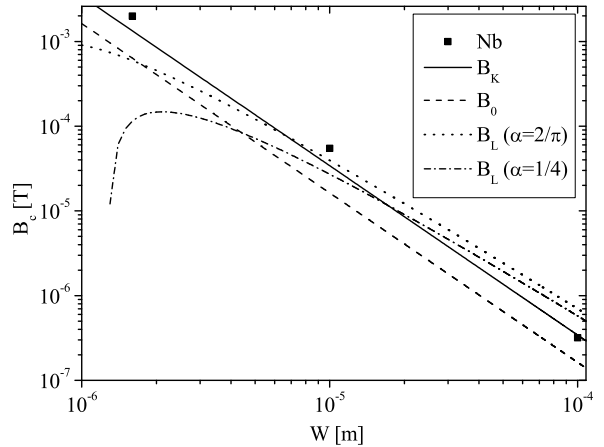}
\caption{\label{fig:martinis} Comparison of experimental results on the critical induction for vortex
exclusion in thin film niobium strips \cite{7890333} with various theories, labeled as in Fig.
\ref{fig:bc}.}
\end{figure}

We also compare results of the present model with previous work on Nb strips
by Stan et al. \cite{7890333} using SHPM. This paper reported critical inductions for 3 different strip widths:
1.6, 10 and 100 $\mu$m. The critical inductions have been compared
to the various models in Fig.\ \ref{fig:martinis}.
The depinning temperature of $T/T_c = 0.9985$ used in this figure was
experimentally determined.\cite{7890333}
Using $\xi_{Nb}(0)=38.9$ nm results in the value $\xi_{Nb}(T=T_{dp}) = 320$ nm  used
for the $B_L$ curves in Fig.\ \ref{fig:martinis}. A reasonably
good agreement exists between the measurements and the predictions of the present model.

\subsection{Trapped vortex density as a function of applied induction}

\begin{figure}[h!]
\includegraphics[width=9cm]{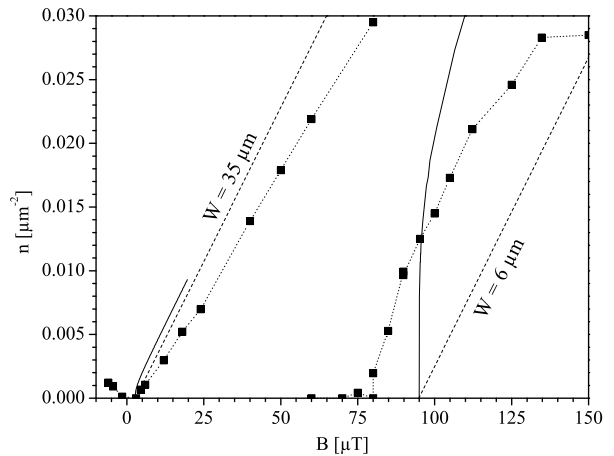}
\caption{\label{fig:nvsb} Plot of the number density of vortices trapped in YBCO strips 35 $\mu$m and 6 $\mu$m
wide as a function of magnetic induction (dots). The dashed lines are the predictions of Eq.\ (\ref{numv}),
without any adjustable parameters. The solid lines indicate the predictions of Eqs.\ (\ref{neff}) and (\ref{smalln})  }
\end{figure}

In Fig.\ \ref{fig:nvsb} the experimentally determined density of trapped vortices
as a function of induction for two strip widths is displayed. This density depends nearly linearly
on the difference between the induction and the critical induction, with a slope
nearly $\Phi_0^{-1}$, in agreement with previous work on Nb strips by Stan et al.\cite{7890333}
The 35 $\mu$m strip width data can be fit to a linear dependence
of the vortex density $n$ on $B_a$ with a slope of
$(3.86\pm0.08)\times 10^{14} ($Tm$^2)^{-1} = (0.83 \pm 0.02)\Phi_0^{-1}$,
with an intercept of $3.8 \pm 1.3\mu$T. The dashed and solid lines in Fig. \ref{fig:nvsb} are the prediction of
the present model [Eqs.\ (\ref{numv}) and (\ref{smalln}) respectively] without any fitting parameters.
Reasonable agreement exists between the present model and measurements.
In the case of the 6 $\mu m$ strips, there is an apparent saturation in the vortex density
for inductions higher than 130 $\mu$T. This may, however be an artifact due to the finite
resolution of our SQUID sensor.
The direction of the applied induction was reversed for three points in the $W=35$ $\mu$m strip data
to check for an offset in the applied induction. Such an offset, if present, was small, as
indicated by the symmetry of the data around zero induction.

\subsection{Vortex spatial distribution}

The local minimum in the Gibbs free energy at $W/2$ of Eq.\ (\ref{Genergy}) makes it
energetically favorable for vortices to be trapped in the center of the strip. However, as the vortex
density increases, the vortex-vortex repulsive interaction makes it
energetically more favorable to form an Abrikosov-like triangular pattern.
Simulations on the trapped vortex position in strips was described by Bronson et al.\cite{8909005}
In particular they predict that there should be a single line of vortices for inductions
$B_c < B_a < 2.48B_c$. Above this induction range a second line of vortices is predicted to form.
As the induction is increased further additional lines of vortices are predicted to form into a nearly
triangular lattice.

\begin{figure}
\includegraphics[width=9cm]{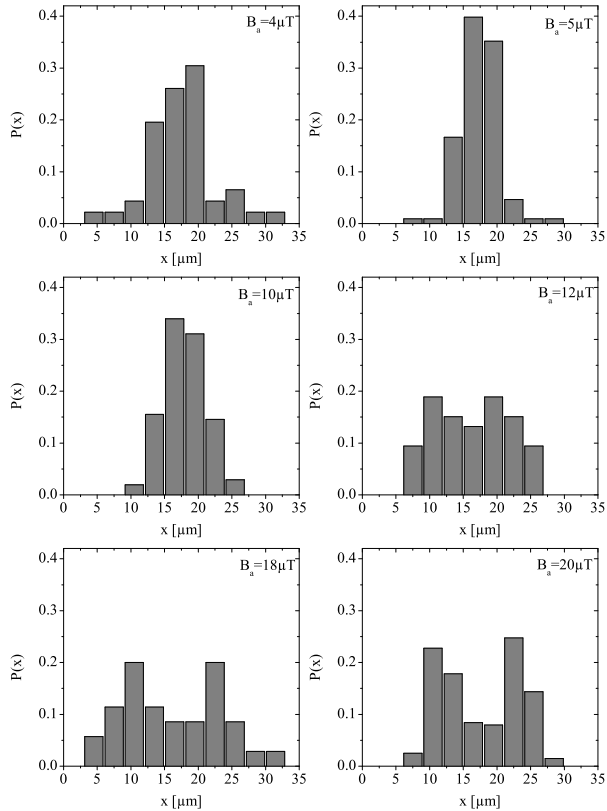}
\caption{\label{fig:lathist} Histograms of the probability of trapping as a function of the
lateral vortex position in a 35 $\mu$m wide YBCO
strip at various inductions. At low inductions the vortices trapped in a single row near
the center of the
film, but above a induction of about 10 $\mu$T they started to reorder.
At a induction of 18 $\mu$T
the vortices were trapped in two relatively well defined rows.}
\end{figure}

We have investigated the distribution of vortices trapped in our strips at various inductions.
As can be seen from the images of Fig.\ \ref{fig:SSM}, even though
there was significant disorder in the vortex trapping positions, there was also some apparent correlation
between the vortex positions. An example can be seen in Fig.\ \ref{fig:lathist}, where a histogram is displayed of the lateral positions of vortices trapped in the 35 $\mu$m wide
strip for several inductions. At low inductions, the vortex lateral position distribution
peaked near the center of the film because the vortices were aligned nearly in a single row.
At a second critical induction of $B_{c2} =11 \pm 1 \mu$T
the distribution started to become broader. At 18 $\mu$T there were
two clear peaks in the distribution, corresponding to two rows. Using the value of $B_c = 3.8 \pm 1.3\mu$T for
the critical induction of the 35 $\mu$m wide strips from our linear fit of the vortex density
vs. applied induction curve of Fig.\ \ref{fig:nvsb}, we found $B_{c2} = (2.89 +1.91-0.93) B_c$. This is
consistent with the prediction of $B_{c2} = 2.48 B_c$ of Bronson et al.\cite{8909005}
In the same paper the critical induction  for the transition from the two-row to the
three-row regime is given to be $B_{c3} = 4.94 B_c$.
This gives $B_{c3} = 18.77 \pm 6.42 \mu$T using the same value for $B_c$.
In our measurements we saw no
evidence for a three row regime.
It was not possible to
perform analysis at higher fields than reported here
because of limitations to the spatial resolution of the SSM.

\begin{figure}
\includegraphics[width=9cm]{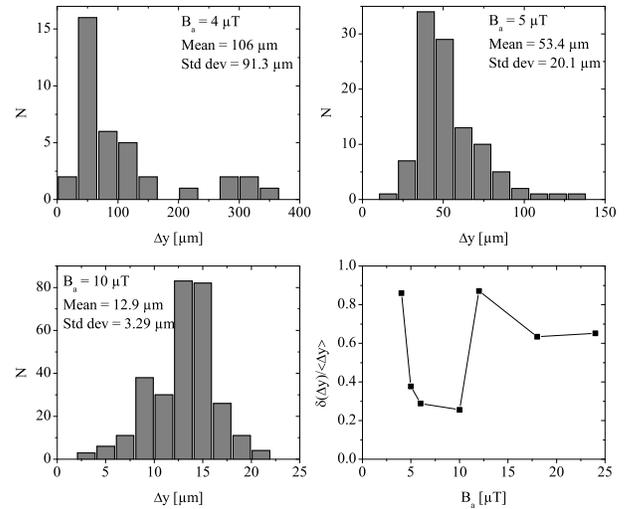}
\caption{\label{fig:longhist} (a-c) Histograms of the longitudinal spacing
between vortices trapped in a 35 $\mu$m wide YBCO strip for selected inductions.
(d) Plot of the standard deviation of the distribution of longitudinal spacings, divided by the
mean of this distribution, as a function of induction. The relative widths of the distributions
became narrower as the induction increases, indicative of ordering in a single row, until at a critical
induction of about 10 $\mu$T there was an abrupt increase in the relative width as two rows started to form.}
\end{figure}

We also saw evidence for longitudinal ordering.
In Fig.\ \ref{fig:longhist}a-c histograms are displayed of the longitudinal distances $\Delta y$
between vortices in the 35 $\mu$m wide strip for various inductions. As expected the
inter-vortex spacing distributions became narrower as the inductions increased, since
the vortex mean spacings decreased. However
the distributions became
narrow faster than their means as the induction was increased, indicative of longitudinal
ordering, until the second critical
induction $B_{c2}$ of approximately 10 $\mu$T was reached. At that induction the relative
distribution width $\delta(\Delta y)/\!\!<\!\!\Delta y\!\!>$ has a discontinuous jump as a second row starts
to form.
A similar decrease in the relative longitudinal distribution width
with increasing induction is observed in the 6 $\mu$m wide strip, although the spatial
resolution of the SSM was not sufficient to resolve vortices at the second critical induction for this width.

In theory there should be longitudinal ordering independent of the magnetic induction.
After all, the Gibbs free energy is independent of the position along the strip and the
only interaction that plays a role is the interaction between the vortices.
Differences in longitudinal ordering as a function of the magnetic induction could arise from
local minima of the Gibbs free energy caused for example by defects in the material.
For relatively low inductions vortices can easily be trapped in defects since the interaction
between the vortices is small because the separation between the vortices is large.
For higher magnetic inductions the number of vortices and likewise the
interaction between the vortices increases. This could mean that the vortices are more
likely to trap at positions determined by the minimization of the vortex-vortex energy
than at positions determined by local defects.

\section{Conclusions}

Experiments on vortex trapping in narrow YBCO strips using a scanning SQUID microscope, as well
as previous measurements on Nb,\cite{7890333} showed a
critical induction for the onset of trapping and a dependence of the vortex density on the
induction which were in good agreement with a new model which takes into account the energy required
to generate a vortex-antivortex pair. In addition, at low inductions the vortices
formed a single row, with longitudinal ordering as the inductions increased. Formation
of a second row was observed at a second critical induction consistent with numerical
modeling.

\section{Acknowledgments}

This research was financed by the Dutch MicroNED program and a VIDI grant (H.H) from the Dutch NWO Foundation. J.R.K was supported by the Center for Probing the Nanoscale, an NSF NSEC, NSF Grant No. PHY-0425897, and by the Dutch NWO Foundation. J.R.C's work at the Ames Laboratory was supported by the Department of Energy-Basic Energy Sciences under Contract No.\ DE-AC02-07CH11358. The SSM setup used in this research was donated to the University of Twente by the IBM T.J.\ Watson Research Center.

\end{document}